\documentstyle [12pt,amsfonts] {article}
\input epsf

\newcommand{\beq}{\begin{equation}}
\newcommand{\eeq}{\end{equation}}
\newcommand{\beqs}{\begin{eqnarray}}
\newcommand{\eeqs}{\end{eqnarray}}

\catcode`@=11
\@addtoreset{equation}{section}
\@addtoreset{equation}{subsection}
\def\theequation{\ifnum\value{section}=0 \arabic{equation}\ignorespaces
\else \ifnum\value{section}=-1 A.\arabic{equation}\ignorespaces
\else \ifnum\value{subsection}=0 \thesection.\arabic{equation}\ignorespaces
\else \thesection.\arabic{subsection}.\arabic{equation}\ignorespaces
                           \fi
                      \fi
                 \fi}
\catcode`@=12

\begin{document}

\def\thefootnote{\fnsymbol{footnote}}

\baselineskip 6.0mm

\vspace{4mm}

\begin{center}

{\Large \bf $T=0$ Partition Functions for Potts Antiferromagnets on M\"obius 
Strips and Effects of Graph Topology} 

\vspace{8mm}

\setcounter{footnote}{0}
Robert Shrock\footnote{email: robert.shrock@sunysb.edu}

\vspace{4mm}

Institute for Theoretical Physics \\
State University of New York       \\
Stony Brook, N. Y. 11794-3840  \\

\vspace{4mm}

{\bf Abstract}
\end{center}

We present exact calculations of the zero-temperature partition function of the
$q$-state Potts antiferromagnet (equivalently the chromatic polynomial) for
M\"obius strips, with width $L_y=2$ or 3, of regular lattices and homeomorphic
expansions thereof.  These are compared with the corresponding partition
functions for strip graphs with (untwisted) periodic longitudinal boundary
conditions.

\vspace{16mm}

\pagestyle{empty}
\newpage

\pagestyle{plain}
\pagenumbering{arabic}
\renewcommand{\thefootnote}{\arabic{footnote}}
\setcounter{footnote}{0}

The chromatic polynomial $P(G,q)$ counts the number of ways that one can color
a graph $G$ with $q$ colors such that no two adjacent vertices have the same
color \cite{birk} (for reviews, see \cite{rtrev}).  The minimum number of
colors needed for this coloring of $G$ is the chromatic number, $\chi(G)$.
Besides its intrinsic mathematical interest, the chromatic polynomial has an
important connection with statistical mechanics since it is the
zero-temperature partition function of the $q$-state Potts antiferromagnet (AF)
\cite{potts,wurev} on $G$: $P(G,q)=Z(G,q,T=0)_{PAF}$.  The Potts AF exhibits
nonzero ground-state entropy $S_0 \ne 0$ (without frustration) for sufficiently
large $q$ on a given lattice graph and is thus an exception to the third law of
thermodynamics (e.g.  \cite{cw}). This is equivalent to a ground state
degeneracy per site $W > 1$, since $S_0 = k_B \ln W$.  Denoting the number of
vertices of $G$ as $n=v(G)$ and $\{G\}=\lim_{n \to \infty}G$, we
have\footnote{\footnotesize{ At certain special points $q_s$ (typically
$q_s=0,1,.., \chi(G)$), one has the noncommutativity of limits $\lim_{q \to
q_s} \lim_{n \to \infty} P(G,q)^{1/n} \ne \lim_{n \to \infty} \lim_{q \to
q_s}P(G,q)^{1/n}$, and hence it is necessary to specify the order of the limits
in the definition of $W(\{G\},q_s)$ \cite{w}.  As in Ref. \cite{w} and our
other works, we shall use the first order of limits here; this has the
advantage of removing certain isolated discontinuities in $W$.}} 
\beq
W(\{G\},q)=\lim_{n \to \infty} P(G,q)^{1/n}
\label{w}
\eeq
Let us consider strips of regular lattices with variable length and fixed
width (with the longitudinal and transverse directions taken to be the $x$ and
$y$ directions, respectively.)  An obvious and important question concerns the
effect of boundary conditions (BC's) on the $T=0$ partition function. 
We study this effect here via new calculations of exact $T=0$ partition 
functions for arbitrarily
long M\"obius strips of regular lattice graphs and certain homeomorphic
expansions thereof, together with a comparison with the corresponding partition
functions for cyclic strips, i.e., strips with (untwisted) periodic
longitudinal boundary conditions.  These strips all have free transverse
boundary conditions (for a comparison of the effects of free versus periodic 
transverse boundary conditions, see \cite{w2d}).  The different longitudinal 
boundary conditions also entail different topology: the cyclic strips 
(topologically $\sim$ cylinder) are oriented, while the M\"obius strips are 
unoriented.  We use the symbols FBC$_x$, PBC$_x = cyc.$, and TPBC$_x=Mb.$ for 
the free, periodic, cyclic, and M\"obius longitudinal boundary conditions, 
respectively.   Some related works on calculations of chromatic polynomials for
families of graphs are Refs. \cite{bds}-\cite{wagner}; see also the list in 
\cite{chia}. 

Since $P(G,q)$ is a polynomial, one can generalize $q$ from ${\mathbb Z}_+$ to
${\mathbb C}$.  
The zeros of $P(G,q)$ in the complex $q$ plane are called chromatic
zeros. Their accumulation set in the limit $n \to \infty$, denoted ${\cal B}$,
is the continuous locus of points where $W(\{G\},q)$ is nonanalytic 
\cite{bkw,read91,w}, \cite{wa}-\cite{nec}. 
The maximal region in the complex $q$ plane to which one
can analytically continue the function $W(\{G\},q)$ from physical values where
there is nonzero ground state entropy is denoted $R_1$.  The maximal value of
$q$ where ${\cal B}$ crosses or intersects with the (positive) real axis is
denoted $q_c(\{G\})$.

We label a particular type of strip graph as $G_s$ and the specific graph of
length $L_x=m$ repeated subgraph units, e.g. columns of squares in the case of
the square strip and hexagons in the case of the kagom\'e strip, as $(G_s)_m$.
If one thinks of the graph as embedded on a rectangular 
strip of paper, with its longitudinal ends glued with the same or
opposite orientation, $L_x$ is the length of this strip in subgraph units.
(Note that, e.g., for the M\"obius square strip, if one starts at a vertex on 
a horizontal boundary edge and traverses a path in the longitudinal 
direction, the pathlength required to return to the original vertex is 
$2L_x$.)  For $m$ greater than a minimal value\footnote{
\footnotesize{For example, for $m \le 4$  ($m \le 3$) the $L_y=2$ cyclic
(M\"obius) square strips exhibit special behavior regarding $g$ and $k_g$;
for $m$ larger than these respective values, they both have $g=4$ and 
$k_g=m$.}}, the cyclic and M\"obius strips of a given $(G_s)_m$ have the same 
number of vertices $n$, edges (bonds) $e$, girth $g$ (length of minimum
closed circuit on $G_s$) and number $k_g$ of circuits of length $g$.  One has 
$n=t_s m$ where $t_s$ depends on $G_s$. Writing 
\beq
P((G_s),q) = \sum_{j=0}^{n-1}(-1)^j h_{n-j}q^{n-j}
\label{p}
\eeq
and using the results that \cite{rtrev,m} 
$h_{n-j}={e \choose j}$ for $0 \le j < g-1$ (whence $h_n=1$ and $h_{n-1}=e$) 
and $h_{n-(g-1)}={e \choose g-1}-k_g$, it follows that for $m$ greater than the
above-mentioned minimal value, these $h_j$'s are the same for the cyclic and 
M\"obius strips of each type $G_s$.  For a given $G_s$, as $m$ increases, the 
$h_{n-j}$'s for the cyclic and M\"obius strips become equal for larger $j$ (see
below). 

A generic form for chromatic polynomials for recursively defined families of 
graphs, of which strip graphs $G_s$ are special cases, is
\beq
P(G_s,q) =  \sum_{j=1}^{N_a} c_j(q)(a_j(q))^m
\label{pgsum}
\eeq 
where $c_j(q)$ and $a_j(q)$ depend on the type of strip graph $G_s$ but are
independent of $m$.  Note that some $c_j$ and $a_j$ may be nonpolynomial
algebraic functions of $q$, in which case certain cancellations occur to yield
the polynomial $P$ \cite{strip,hs}. 
The simplest cyclic strip is the circuit graph with $m$
vertices, $C_m$, for which $P(C_m,q)=(q-1)^m+(q-1)(-1)^m$. The chromatic
polynomials for the cyclic and twisted cyclic ladder graphs were calculated in
\cite{bds} and have $N_a=4$. $W$ and ${\cal B}$  were given for these and 
other families in \cite{w} (see also \cite{read91}).  Various iterative 
methods have been used in the past to calculate chromatic polynomials for 
recursive families of graphs 
\cite{bds}, \cite{bkw}-\cite{read91}, \cite{w}, \cite{wa}-\cite{bn}.
\footnote{\footnotesize{The method used in \cite{b,bn} can also be 
used to obtain rigorous bounds on $W$ for 2D lattices \cite{b,ww,wn}.}}
A method that we have found useful to compute $P((G_s)_m,q)$ for a variety of
strips with FBC$_x$ \cite{strip,hs} and (T)PBC$_x$ \cite{pg,wcy,nec} makes 
use of a generating function
\beq
\Gamma(G_s,q,x) = \sum_{m=m_0}^{\infty}P((G_s)_m,q)x^{m-m_0}
\label{gamma}
\eeq
where $\Gamma(G_s,q,x)={\cal N}(G_s,q,x)/{\cal D}(G_s,q,x)$ with 
\beq
{\cal N}(G_s,q,x) = \sum_{j=0}^{deg_x{\cal N}} A_j(q) x^j
\label{n}
\eeq
and
\beq
{\cal D}(G_s,q,x) =  1 + \sum_{j=1}^{deg_x{\cal D}} b_j(q) x^j = 
\prod_{j=1}^{deg_x{\cal D}}(1-\lambda_j(q)x)
\label{d}
\eeq
where $m_0$, $A_j$, and $\lambda_j$ depend on the type of strip, and $A_j$ and
$b_j$ are polynomials in $q$ ($m_0=2$ for the square and homeomorphically
expanded square strips and $m_0=1$ for the kagom\'e strips considered below).  
$\Gamma$ is calculated by means of standard
theorems for chromatic polynomials such as the deletion-contraction theorem. 
The $a_j$ and $c_j$ in (\ref{pgsum}) can be determined from $\Gamma$ using
eqs. (2.17)-(2.21) of \cite{hs} (taking into account value of $m_0$ for a given
strip).  For a specified $G_s$, the $\lambda$'s in 
${\cal D}$ that contribute to $P$ are identical to the\footnote{
\footnotesize{For our square lattice strips, all $\lambda_j$'s
contribute to $P$.  For the kagom\'e strip and homeomorphically expanded square
strips considered here, there are, respectvely, three $\lambda$'s and one
$\lambda$ that do not contribute to $P$ \cite{wcy}.}} $a_j$'s.
For a given type of strip, we shall study the effect of cyclic versus M\"obius
longitudinal boundary conditions on (i) the set of $a_j$, (ii) the $W$ 
function; (iii) the locus ${\cal B}$, (iv) the set of $c_j$, and (v) $\chi$.  
Quantities (i)-(iv) depend only on
$G_s$ while (v) also depends on $m$. \ \footnote{\footnotesize{It is also of
interest to compare chromatic polynomials for lattice strips with 
cyclic versus free longitudinal boundary conditions; see 
\cite{strip,hs}, \cite{pg}-\cite{nec}.}}  

We consider first strips of the square (sq) lattice of length $L_x=m \ge 2$ 
and width $L_y$ vertices.  The $L_y=1,2$ cases have been mentioned above. 
For the cyclic $L_y=3$ strip, $P$, $W$, and ${\cal B}$ were calculated in 
\cite{wcy}.  Our new result for the $L_y=3$ M\"obius strip is 
\beqs 
& & P(sq(L_y = 3,\ Mb.)_m,q) = (q^2-3q+1)(-1)^m
-(q-1)^m+(q-2)^m-(q-4)^m \cr\cr & & -(q-1)[-(q-2)^2]^m + \Bigl [
(\lambda_{sq,6})^m+(\lambda_{sq,7})^m \Bigr ] +(q-1)\Bigl
[(\lambda_{sq,8})^m+(\lambda_{sq,9})^m+(\lambda_{sq,10})^m \Bigr ] \cr\cr &&
\label{psqly3tw}
\eeqs
where
\beq
\lambda_{sq,(6,7)} = \frac{1}{2}\Biggl [ (q-2)(q^2-3q+5) \pm 
\Bigl \{ (q^2-5q+7)(q^4-5q^3+11q^2-12q+8) \Bigr \}^{1/2} \Biggr ]
\label{lambda67}
\eeq
and $\lambda_{sq,j}$, $j=8,9,10$, are the roots of the cubic equation
\beq
\xi^3+b_{sq,21}\xi^2+b_{sq,22}\xi+b_{sq,23}=0
\label{sqcubic}
\eeq
with 
\beq
b_{sq,21}=2q^2-9q+12
\label{bsq21}
\eeq
\beq
b_{sq,22}=q^4-10q^3+36q^2-56q+31
\eeq
\label{bsq22}
\beq
b_{sq,23}=-(q-1)(q^4-9q^3+29q^2-40q+22) \ .
\label{bsq23}
\eeq

For comparison, the cyclic $L_y=3$ result is \cite{wcy}
\beqs
& & P(sq(L_y = 3,cyc.)_m,q) = (q^3-5q^2+6q-1)(-1)^m \cr\cr
& & + (q^2-3q+1)\Bigl [(q-1)^m+(q-2)^m+(q-4)^m \Bigr ]
+ (q-1)[-(q-2)^2]^m \cr\cr
& & + \Bigl [ (\lambda_{sq,6})^m+(\lambda_{sq,7})^m \Bigr ]
+(q-1)\Bigl [(\lambda_{sq,8})^m+(\lambda_{sq,9})^m+(\lambda_{sq,10})^m \Bigr ]
\ .
\label{psqly3}
\eeqs

We discuss these further below.

We next consider strips of the kagom\'e ($kg$) lattice comprised of $m$
hexagons with each adjacent pair sharing two triangles as adjacent polygons. 
For the M\"obius kagom\'e strip with $L_y=2$ we calculate
\beqs
& & P(kg(L_y=2, \ Mb.)_m,q) = -(q-4)^m -(q-1)\Bigl [(q-1)(q-2)^2 \Bigr ]^m 
\cr\cr 
& & + \Bigl [(\lambda_{kg,1})^m+(\lambda_{kg,2})^m \Bigr ] + 
(q-1)\Bigl [(\lambda_{kg,3})^m+(\lambda_{kg,4})^m \Bigr ]
\label{pkagtw}
\eeqs
where 
\beqs
& & \lambda_{kg,(1,2)} = \frac{1}{2} \biggl [ q^4-6q^3+14q^2-16q+10 \cr\cr
& & \pm \Bigl \{q^8-12q^7+64q^6-200q^5+404q^4-548q^3+500q^2-292q+92
  \Bigr \}^{1/2} \biggr ]
\label{lamkg12}
\eeqs
\beq
\lambda_{kg,(3,4)} = \frac{1}{2} \biggl [q^3-7q^2+19q-20  
\pm \Bigl \{q^6-14q^5+83q^4-278q^3+569q^2-680q+368 \Bigr \}^{1/2} \biggr ] \ .
\label{lamkg34}
\eeq

In contrast, the chromatic polynomial of the regular cyclic $L_y=2$ kagom\'e
strip is \cite{wcy}
 \beqs
& & P(kg(L_y=2,cyc.)_m,q) =
(q^2-3q+1)(q-4)^m+(q-1)\Bigl [(q-1)(q-2)^2 \Bigr ]^m \cr\cr & & + \Bigl
[(\lambda_{kg,1})^m+(\lambda_{kg,2})^m \Bigr ] + (q-1)\Bigl
[(\lambda_{kg,3})^m+(\lambda_{kg,4})^m \Bigr ] \ . 
\label{pkag}
\eeqs

It is also of interest to study homeomorphic expansions of the cyclic square
strip.\footnote{\footnotesize{ A homeomorphic expansion of a graph consists in
the addition of degree-2 vertices to bonds of that graph.}}  We start with the
cyclic or M\"obius $L_y=2$ square strip and add $k-2$ vertices to each upper
bond (edge).  Define 
\beq
D_k = \frac{P(C_k,q)}{q(q-1)} = 
\sum_{s=0}^{k-2}(-1)^s {{k-1}\choose {s}} q^{k-2-s}
\label{dk}
\eeq
where $P(C_k,q)=(q-1)^k+(q-1)(-1)^k$ is the chromatic polynomial for the
circuit graph with $k$ vertices.  For the twisted cyclic strip we calculate 
\beq 
P((Ch)_{(k,2),Mb.,m},q) = -(-1)^{km} + (D_{k+2})^m +
\Bigl ( \frac{D_{k+1}+(-1)^kq}{\lambda_{h,4}-\lambda_{h,5}} \Bigr ) \Bigl
[(\lambda_{h,4})^m-(\lambda_{h,5})^m \Bigr ]
\label{phet}
\eeq
where
\beq
\lambda_{h,(4,5)}=\frac{1}{2}\Biggl [ -D_{k+1}-(-1)^k(q-2)\pm \Bigl \{
[D_{k+1}-(-1)^k(q-2)]^2+4D_k(-1)^k \Bigr \}^{1/2} \ \Biggr ] \ , 
\label{lamhom45}
\eeq
$h$ means homeomorphic, and $(k_1,k_2)$ refers to the respective
number of vertices (counting the end ones) on the upper and lower horizontal
bounds between each vertical bond.

For the cyclic case \cite{wcy}, 
\beq
P((Ch)_{(k,2),cyc.,m},q) = (q^2-3q+1)(-1)^{km} +
(D_{k+2})^m + (q-1)\Bigl [ (\lambda_{h,4})^m+(\lambda_{h,5})^m \Bigr ] \ . 
\label{phe}
\eeq
For the symmetric homeomorphic expansion, $P((Ch)_{(k,k),cyc.,m},q)$ and 
$P((Ch)_{(k,k),Mb.,m},q)$ and the resultant $W$ and ${\cal B}$ were given in 
\cite{pg}.  Results for $P$, $W$, and ${\cal B}$ for the cyclic and M\"obius 
$L_y=2$ triangular strips were given in \cite{wcy} and $P$ for the cyclic 
$L_y=2$ triangular strips also in \cite{bn}.  Note that for
the $L_y=2$ M\"obius triangular strip and homeomorphically expanded
square strip in eq. (\ref{phet}) two of the $c_j$'s are nonpolynomial 
algebraic functions of $q$. 

We comment on how the nonpolynomial algebraic roots in the various $P$'s yield
polynomials in $q$.  As is evident from (\ref{d}), for a given strip, the
$\lambda_j$'s arise as roots of the equation ${\cal D}=0$.  In general, 
${\cal D}$ contains some number of factors of linear, quadratic, cubic, etc. 
order in $x$.  Consider a generic factor in ${\cal D}$ of $r$'th degree in 
$x$: \ $(1+f_1x+f_2x^2+...+f_rx^r)$, where the $f_j$'s are polynomials in 
$x$.  This yields $r$ \ $\lambda_\ell$'s as roots of the equation 
$\xi^r+f_1\xi^{r-1}+...+f_r=0$. The expressions in $P$ involving these roots 
are symmetric polynomial functions of the roots, namely terms of the form 
\beq
s_m = \sum_{\ell=1}^r (\lambda_\ell)^m
\label{s}
\eeq
and, for certain M\"obius strips (as well as strips with FBC$_x$), terms of 
the form 
\beq
\frac{(\lambda_a)^m-(\lambda_b)^m}{\lambda_a-\lambda_b} = 
\sum_{k=0}^{m-1}(\lambda_a)^{m-1-k}(\lambda_b)^k \ . 
\label{ldif}
\eeq
We can then apply a standard theorem (e.g. \cite{uspensky}) which states that 
a symmetric polynomial function of the roots of an algebraic equation is a 
polynomial in the coefficients, here $f_\ell$, $\ell=1,..,r$; hence this 
function is a polynomial in $q$.  For example, for $s_r$, one has the 
well-known formulas (due to Newton) $s_1=-f_1$, $s_2=f_1^2-2f_2$, 
$s_3=-f_1^3+3f_1f_2-3f_3$, etc. 

Concerning question (i), in general we find that for a given type of lattice
strip (or homeomorphic expansion thereof), the terms $a_j$ in (\ref{pgsum}) are
the same for the cyclic and M\"obius strips.  This can understood as
follows. Recall that the linear equation for the elongation of the strip,
together with corresponding equations for related strips produced via
deletion-contraction operations, form the matrix equation $T v_{_\Gamma} = v_0$
(eq. (2.29) of Ref.  \cite{strip}), where $v_{_\Gamma}$ is the vector comprised
by the generating function $\Gamma$ for the strip of interest, together with
generating functions for the above-mentioned related strips, and $v_0$ is a
vector of chromatic polynomials for beginning forms of these strips.  The
denominator ${\cal D}$ of the generating function $\Gamma$ is then given as
${\cal D} = det(T)$.  The set of generating functions in the vector
$v_{_\Gamma}$ does depend on the longitudinal boundary conditions, as is clear
from the fact that it has larger dimension for periodic than free BC$_x$'s.
However, it is the same for cyclic and M\"obius BC$_x$'s 
(the elongation is a local operation); the difference here is in the
chromatic polynomials in $v_0$.  For example, in the
simplest case of the $L_y=2$ square strip, the minimal subgraphs, viz., for
$m=2$, are a single square, or equivalently, circuit graph $C_4$ for PBC$_x$
but the complete graph $K_4$ for TPBC$_x$ (M\"obius).\footnote{The complete
graph $K_p$ is the graph consisting of $p$ vertices all adjacent to each
other.}  Hence, ${\cal D}$ is the same, while ${\cal N}$ is different, for a
given strip with cyclic and M\"obius longitudinal BC's.  Since the
$\lambda_j$'s in ${\cal D}$ that contribute to $P$ are identical to the $a_j$'s
in eq. (\ref{pgsum}) (cf. eq. (2.20) in \cite{hs}), this enables one to
understand that these $a_j$'s are the same for the cyclic and M\"obius strips
of a particular type, given the observed property that in the cases where
certain $\lambda_j$'s do not contribute to $P$, these are the same for the 
cyclic and M\"obius cases. 

The property that the $a_j$'s are the same for these two cases also answers 
questions (ii) and (iii).  The term
$a_j$ that has maximal magnitude in a region of the $q$ plane determines $W$ in
that region, and hence the locus ${\cal B}$ is the solution of the equation
$|a_j|=|a_{j'}|$ of terms of maximal magnitude.  It follows that $W$ and ${\cal
B}$ are the same for the cyclic and M\"obius versions of a particular kind of
strip.  The definition (\ref{w}) and the equality of the $W$ functions for 
the cyclic and M\"obius strips of a given type $G_s$ imply that, as noted
above, as $n$ increases, the $h_{n-j}$'s for these two strips become equal for
progressively larger values of $j$. 

On question (iv), since the $c_j$ are functions of both the $\lambda_j$'s in
${\cal D}$ and the coefficient functions $A_j$ in ${\cal N}$, and since the
latter differ for cyclic and M\"obius BC$_x$'s, the $c_j$'s will also, in
general, differ.  However, a subset may be the same.  In particular, 
the coefficient $c_j$ of the term which is dominant
in region $R_1$ is always unity, as a consequence of the property that $h_n=1$
in eq. (\ref{p}).  Hence, {\it a fortiori}, this coefficient is the same for
cyclic and M\"obius BC$_x$'s.  In cases where the leading $a_j$ is a
nonpolynomial algebraic function of $q$, the requirement that $P$ be a
polynomial implies that certain other coefficients must also be unity.  Thus,
for example, $c_6$ and $c_7$ are both equal to 1 for the $L_y=3$ square strip,
and the $(\lambda_{kg,1})^m$ and $(\lambda_{kg,2})^m$ terms both have 
coefficient 1 for the kagom\'e strip. 

The situation with the chromatic number, question (v), depends on the
particular strip.  In the following, we take $L_x=m$ greater than the
above-mentioned minimal values where the strips degenerate.  
The cyclic square strips have $\chi=2$ for $L_x$ even and 
$\chi=3$ for $L_x$ odd, independent of $L_y$.  For M\"obius square strips, 
$\chi=2$ for $(L_x,L_y)=(e,o)$ or $(o,e)$, and $\chi=3$ for
$(L_x,L_y)=(e,e)$ or $(o,o)$, where $e$ and $o$ denote even and odd.  Both the
cyclic and M\"obius kagom\'e strips above have $\chi=3$, which shows that the
twisting does not necessarily change $\chi$.  The cyclic and M\"obius
homeomorphically expanded square strips [eqs. (\ref{phe}) and (\ref{phet})] 
both have $\chi=3$ for odd $k$, independent of $m$, while for even $k$, the 
cyclic strip has $\chi=2, \ 3$ for $m$ even, odd, and the M\"obius strip
has $\chi=2, \ 3$ for $m$ odd, even, respectively. 

For a given type of strip $G_s$, define $C=\sum_{j=1}^{N_a}c_j$.  For 
the $L_y=2$ square strips, $C=q(q-1)$ and 0 for the cyclic and M\"obius cases
\cite{bds,bn}.
  This is also true for the $L_y=2$ cyclic and M\"obius 
triangular strips, respectively, as also for the cyclic and twisted 
homeomorphically expanded $L_y=2$ square strips.   For the $L_y=3$ cyclic
square strip, eq. (\ref{psqly3}) yields $C=q(q-1)^2$, in accord with the
generalization $C=P(T_{r=L_y},q)=q(q-1)^{L_y-1}$ for the cyclic square strip of
width $L_y$. For the $L_y=3$ square M\"obius strip, eq. ({\ref{psqly3tw}) gives
$C=q(q-1)$.  The $L_y=2$ cyclic and M\"obius kagom\'e strips have $C=q^2$ and 
$C=q$, respectively. 

Our comparative study thus elucidates the effect of periodic versus twisted
periodic longitudinal boundary conditions on the $T=0$ Potts antiferromagnet
$T=0$ partition function (chromatic polynomial) for families of lattice strip
graphs and homeomorphic expansions thereof.  In the future it would also be of
interest to extend this study to consider other combinations of longitudinal
and transverse boundary conditions.

\vspace{4mm}

This research was supported in part by the NSF grant PHY-97-22101.

\vfill
\eject
\end{document}